# A Comparative Study of the Valence Electronic Excitations of $N_2$ by Inelastic X-ray and Electron Scattering

J.A. Bradley[1,2], G.T. Seidler[1,*], G. Cooper[3], M. Vos[4], A.P. Hitchcock[3], A. P. Sorini[1,5], C. Schlimmer[1], K.P. Nagle[1]

*1. Physics Department, University of Washington, Seattle, WA 98195*
*2. Los Alamos National Laboratory, Los Alamos, NM, 87545*
*3. Brockhouse Institute for Materials Research, McMaster University, Hamilton, Ontario, Canada L8S 4M1*
*4. Atomic and Molecular Physics Laboratories, Research School of Physics and Engineering, Australian National University Canberra, Australia*
*5. Stanford Institute for Materials and Energy Sciences, SLAC National Laboratory, Menlo Park, CA, 94025*

Bound state, valence electronic excitation spectra of $N_2$ are probed by nonresonant inelastic x-ray and electron scattering. Within usual theoretical treatments, dynamical structure factors derived from the two probes should be identical. However, we find strong disagreements outside the dipole scattering limit, even at high probe energies. This suggests an unexpectedly important contribution from intra-molecular multiple scattering of the probe electron from core electrons or the nucleus. These effects should grow progressively stronger as the atomic number of the target species increases.

The inelastic scattering of electrons, as in electron energy loss spectroscopy (EELS), provides a valuable tool to study both dipole-allowed and dipole-forbidden electronic excitations. This has proven central to numerous scientific and technical disciplines, including: fundamental molecular physics and chemistry [1, 2], optical properties of terrestrial and planetary atmospheres [3], and numerous branches of energy science [4]. While the simplest theoretical treatments dating to the earliest days of quantum scattering theory have proven to be a valuable starting point in essentially all cases, there is steadily growing evidence that such treatments are often, and perhaps in general, fundamentally inadequate [5-6, 7, 8]. There would be great value in having an alternative experimental technique that probes the same kinematic parameter space as inelastic electron scattering but whose scattering dynamics unequivocally obey the lowest-order theoretical approaches. Non-resonant inelastic X-ray scattering (NRIXS) using hard X-rays provides this approach [9].

In this paper, we compare bound-state valence electron excitations of molecular $N_2$ using inelastic scattering by electrons and by hard x-rays. The electron and photon scattering results disagree upon leaving the low momentum transfer limit, with gross qualitative and quantitative deviations at high momentum transfer. We discuss these results in the context of multiple intra-molecular scattering, wherein the probe electron transfers significant energy and limited momentum to the electronic degrees of freedom along with large momentum and little energy transfer to the core electrons and nucleus of the target species. If this explanation is correct, the observed phenomenon should scale strongly with Z and become progressively more pronounced for heavier elements.

Under the simplest scattering approximations, NRIXS and EELS are equivalent probes of target electronic structure, with double differential cross sections (DDCS) given by

$$\left(\frac{d^2\sigma}{d\Omega d\omega}\right)_{\gamma,e} = \left(\frac{d\sigma}{d\omega}\right)_{Th,Ru} S(\mathbf{q},\omega) , \quad (1)$$

where $\gamma$, $e$ refer to photon and electron scattering, respectively, and $Th$, $Ru$ refer to Thomson and Rutherford differential cross-sections. The variable $q$ refers to momentum transfer, $\omega$ is probe particle energy loss, and $\Omega$ is detection solid angle. The quantity $S(q,\omega)$ is known as the material's *dynamic structure factor*. For inelastic photon scattering (NRIXS), Eq. 1 is derived directly from first order perturbation in the $\vec{A}\cdot\vec{A}$ electron-photon interaction Hamiltonian. Extensive evidence demonstrates that Eq. 1 holds with considerable certainty for NRIXS [9]. For EELS, however, the conclusion relies on a weak probe-target electron interaction (the FBA), the purely binary interaction between the probe electron and the quantum electronic excitations of the target (the binary encounter approximation (BEA)), and the negligible influence of the internal structure of the target on the asymptotic form of the probe wavefunction (the plane wave impulse approximation (PWIA)) [8]. Here, we label the experimentally determined $S(\mathbf{q},\omega)$ as $S_\gamma(q,\omega)$ when it is derived from NRIXS and as $S_e(q,\omega)$ when it is derived from EELS.

To explain the large differences we will demonstrate between $S_\gamma(\mathbf{q},\omega)$ and $S_e(\mathbf{q},\omega)$, it is important to briefly review the expected behavior of $S(\mathbf{q},\omega)$ [10]. To begin,

$$S(\mathbf{q},\omega) = \sum_f \left|\langle u_f | e^{i\mathbf{q}\cdot\mathbf{r}} | u_i \rangle\right|^2 \delta(E_f - E_i - \hbar\omega) \quad (2)$$

where $u_i$ and $u_f$ represent target initial and final electronic states with energies $E_i$ and $E_f$. In the momentum basis,

$$\langle u_f | e^{i\mathbf{q}\cdot\mathbf{r}} | u_i \rangle = \int d^3 p \cdot u_f^*(\mathbf{p}) u_i(\mathbf{p}+\mathbf{q}) \quad (3)$$

where $u_{f,i}(\mathbf{p})$ are the momentum-space final and initial wavefunctions, respectively. Note that the transferred momentum causes a relative shift of the arguments for the momentum-space wavefunctions, but otherwise Eq. 3 is simply an overlap integral. A large enough shift (i.e., momentum transfer in the scattering event) will render the overlap nil. For bound final states, $S(\mathbf{q},\omega)$ should rapidly vanish as $q$ grows large compared to the widths of the momentum-space wavefunctions. Below, we will discuss the explicit dependence of $S(\mathbf{q},\omega)$ on $q$, but the effective valence shell size of $d_{eff} \sim 4$ a.u. leads to a general momentum transfer scale of $2\pi/d_{eff} \sim 1.5$ a.u. above which the aforementioned considerations must eventually dominate.

The dependence of $S(\mathbf{q},\omega)$ on the relevant selection rule for a bound state excitation follows from expanding the exponential operator of Eq. 2 in spherical harmonics and performing a directional average appropriate for polycrystalline or gaseous samples,

$$\langle u_f | e^{i\mathbf{q}\cdot\mathbf{r}} | u_i \rangle = \sum_{l,m} 4\pi i^l \langle u_f | j_l(qr) Y_{l,m}(\hat{\mathbf{r}}) | u_i \rangle. \quad (4)$$

The $q$-dependence is entirely in the spherical Bessel function. Since $j_l(qr) \to 0$ at order $(qr)^l$ as $qr \to 0$, we can make the following general statements: At low $q$, $S(q,\omega)$ shows dipole-allowed transitions exclusively. As $q$ grows, the dipole transitions fade away, and quadrupole transitions dominate $S(q,\omega)$. As $q$ continues to grow, this process repeats at higher and higher multipoles: quadrupole transitions fading and octupole transitions rising, and so on [10]. This behavior is commonly termed $q$-dependent *multipole selection rules*. For N$_2$ gas, we will demonstrate that these selection rules break down for EELS-derived $S_e(q,\omega)$, while they remain intact for NRIXS-derived $S_\gamma(q,\omega)$.

All NRIXS measurements were performed with the lower energy resolution inelastic x-ray scattering (LERIX) spectrometer [11] at sector 20-ID-B of the Advanced Photon Source. To obtain the spectra, incident photon energy was scanned between 9896 and 9910 eV. The beamline monochromator was operated with either a double Si (111) or a double Si (311) configuration for 1.4 or 0.9 eV net energy resolution, respectively. Allowing for the differing energy resolutions in the two studies (1.4 eV versus 0.9 eV), the spectra were mutually consistent. The lowest $q$ (0.42 a.u.) analyzer was misaligned for the high-resolution data collection, so the 1.4 eV resolution data is used. Otherwise, all data are from the 0.9 eV resolution measurements. An NRIXS-compatible gas-phase pressure cell [12] was used to collect data at 1.0 MPa pressure, with a total integration time of 120 sec per incident energy. Spectra were verified to be independent of gas pressure. After correction for known systematic effects, $S_\gamma(q,\omega)$ was converted to units of $(eV \cdot molec)^{-1}$ by application of the Bethe *f*-sum rule [9, 13]. The NRIXS normalization to absolute units is estimated correct to within 10% [14].

EELS measurements were performed with two different instruments. Initial measurements were performed with an EELS spectrometer [15] at McMaster University using an incident electron energy of 2.25 keV. Spectra were verified to be independent of gas pressure. The quantity $S_e(q,\omega)$ was normalized to the published N$_2$ elastic cross-section [16]. The McMaster EELS measurements were performed with 0.7 eV energy resolution, but have been broadened to 0.9 eV final resolution for ease of comparison with NRIXS spectra. Additional measurements were performed with an EELS spectrometer at the Australian National University (ANU) [17]. The incident energies used in the ANU EELS measurements ranged from 0.6 to 6.0 keV and the energy resolution was 0.6 eV.

In Figure 1, we compare $S_\gamma(q,\omega)$ and $S_e(q,\omega)$ for the low-energy (valence) electronic excitations of $N_2$ gas. When the present EELS results are compared to prior studies [18] of the dipole-forbidden, quadrupole-allowed Lyman-Birge-Hopfield (LBH) band at ~9 eV, they agree favorably [14]. Fig. 1 shows that, within the combined uncertainty in the EELS and NRIXS normalizations, the lowest-$q$ EELS and NRIXS data are in agreement. In the limit $q \to 0$, it is well established that both $S_\gamma(q,\omega)$ and $S_e(q,\omega)$ become proportional to the optical absorption spectrum, so this result is expected and it verifies our current methodology.[9, 19]

At higher $q$, however, $S_e(q,\omega)$ and $S_\gamma(q,\omega)$ rapidly diverge from one another. $S_\gamma(q,\omega)$ behaves as expected: First, there is an overall decrease in oscillator strength as $q$ grows larger than the momentum-space width of the valence-state wavefunction. Second, $S_\gamma(q,\omega)$ exhibits the expected multipole selection rules in the $q$-dependent transition intensities. For example, the lowest energy excitation of $N_2$ (the $a^1\Pi_g$ feature) is known to be dipole forbidden and quadrupole allowed, and indeed, for NRIXS at low $q$ this feature is strongly damped, rises for mid-range $q$, and then falls away at high $q$. On the other hand, $S_e(q,\omega)$ displays markedly different $q$ dependence. In the inelastic electron scattering results, spectral features rise up, but they do not fade away at higher $q$

The breakdown of the selection rules, when coupled with the firmer footing of the approximations leading to Eq. 1 in the NRIXS case, is experimental evidence for the failure of one or more of the approximations leading to Eq. 1 for EELS.

The consistency of features in $S_e(q,\omega)$ as $q$ grows has an important consequence: Since the energy transfer is known to be entirely due to the electronic degree of freedom at low $q$, this must extend to high $q$, even though $S_e(q,\omega)$ and $S_\gamma(q,\omega)$ disagree. With this in mind, the unexpectedly high transition amplitudes in $S_e(q,\omega)$ at high q are puzzling. Valence electronic excitations in a single scattering event are not equipped to take up this momentum, even though they are known to take up the energy transfer. This suggests multiple scattering, though multiple inter-molecular scattering is ruled out by the independence of the EELS spectra on $N_2$ gas density. Instead, we propose that the inelastic electron scattering results are due to multiple *intra*-molecular scattering, in which nearly all the energy transfer occurs in a scattering event between the electron probe and the valence electron configuration, but some (possibly large) fraction of the momentum transfer occurs via (quasi)elastic scattering between the probe electron and the target's core electrons or nucleus.

This model is supported by both contemporary work on electron-impact ionization [2, 20] and also by older work on the lower-energy excitations to bound states [5, 7, 21-22]. In the former case, the key experimental observation has been the discovery of a significant momentum transfer to the residual ion, which was most convincingly explained by making corrections to the plane-wave approximation that reflect the gradually decreased screening of the nuclear charge at small classical impact parameters.[2] In the latter case, the early EELS study of He by Opal and Beaty [23] and corresponding distorted-wave Born calculations of Hidalgo and Geltman,[24] which includes a dominant effect from nuclear scattering at high $q$, stand largely alone as a well-documented example of FBA violation for *bound state* excitations in EELS. Similar issues have also been elegantly investigated by Kelsey [7] with the use of a somewhat unconventional second Born approximation — one scattering event from an electronic potential and a second from the nuclear potential. The various theoretical approaches are complementary treatments of the same problem: a direct sensitivity of the probe electron to the detailed internal structure of the target, in violation of some subset of the FBA, PWIA or BEA.

A subtlety arises out of this discussion. While it is well know that the FBA is violated at low incident electron energy in EELS, here we have found evidence for a violation of Eq. 1 at high incident electron energies. Similar considerations arose in some of the earliest treatments leading to distorted-wave Born approximations, and predicted asymptotic violations of the FBA.[21] In Fig. 2, we further explore this by presenting EELS spectra at higher incident energy and higher $q$, obtained with the ANU spectrometer. Note that the 6 keV, 45° and the 1 keV, 135° measurement differ sharply in terms of one spectral feature (~13 eV), but are otherwise similar throughout the rest of this range of energy loss. The momentum transfer in these two measurements is quite close; consequently, the EELS DDCS is not a function of $q$ and $\omega$ only. Furthermore, note that comparison of the McMaster $q = 5.5$ a.u. (Fig. 1) with the ANU $q= 5.1$ a.u. or $q = 6.6$ a.u. (Fig. 2) data alone could easily lead to the erroneous conclusion that Eqn. 1 is satisfied, since the spectra would seem to be independent of incident energy.

In summary, the bound-state electronic excitation spectra for $N_2$ exhibit profound qualitative and quantitative differences when they result from nonresonant inelastic electron as opposed to nonresonant inelastic x-ray scattering, even at very similar momentum transfers. This behavior is consistent with intra-molecular multiple scattering of electrons, giving additional sensitivity to the internal structure of the target and especially to the presence of core-level electrons and the nuclear charge. Such effects are expected to grow progressively stronger with increasing Z in the limit of high momentum transfer,[7] and thus may have far-reaching consequences for inelastic electron scattering from bound state excitations, especially for heavier target species.

We thank Edward Kelsey, Donald Madison, William McCurdy, Charles Malone, and Paul Johnson for useful discussions. This work was supported by the U.S. Department of Energy, the Natural Sciences and Engineering Research Council (NSERC) of Canada, and the Australian Research Council. Measurements at the Advanced Photon Source (APS) are supported by the U.S. Department of Energy, NSERC of Canada, the University of Washington, Simon Fraser University and the APS.

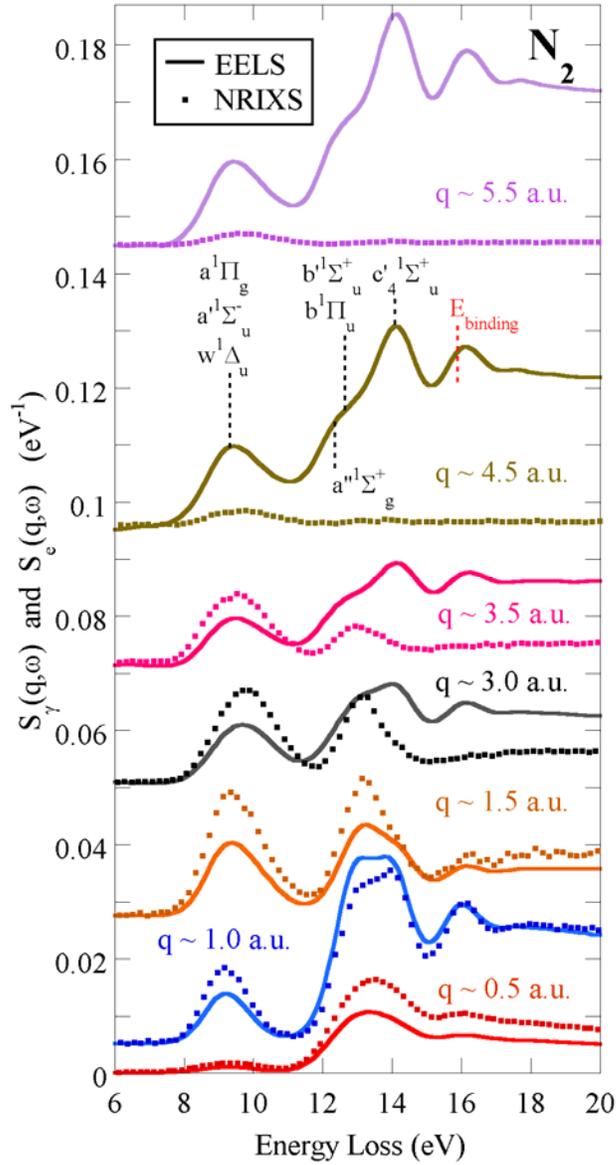

FIG. 1. NRIXS-derived $S_\gamma(q,\omega)$ and EELS-derived $S_e(q,\omega)$ from nitrogen gas with symmetry designations labeled for bound state final state. Values of $q$ for EELS and NRIXS spectra are within 0.2 a.u. for all $q$'s listed, with the exception of the q=1.5 a.u. curve, where NRIXS $q$ is 1.23 a.u. and EELS $q$ is 1.58 a.u.

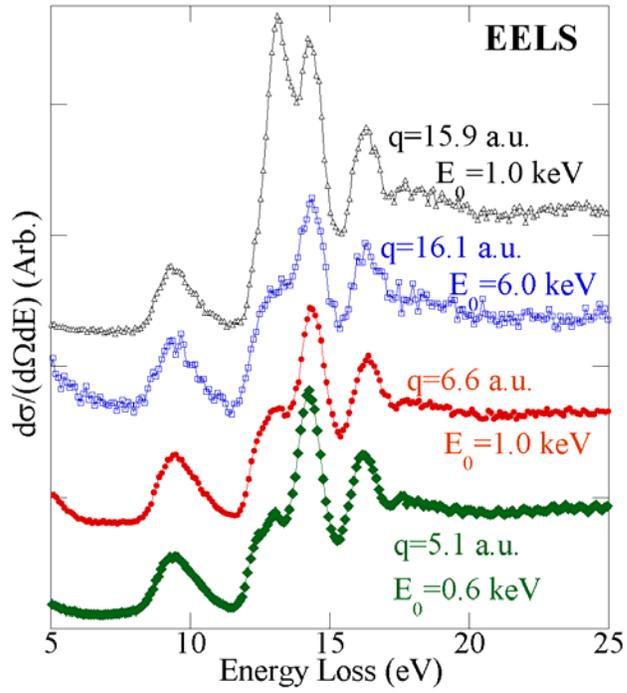

FIG. 2. The $N_2$ EELS scattering cross section at various incident electron energies ($E_0$), and q-values. For the top-most measurement the scattering angle was 135°, while it was 45° for the others.